\documentclass[letterpaper, 10 pt, conference]{ieeeconf}  

\IEEEoverridecommandlockouts                              

\usepackage[utf8]{inputenc} 
\usepackage[T1]{fontenc}    
\usepackage{hyperref}       
\usepackage{url}            
\usepackage{booktabs}       
\usepackage{amsfonts}       
\usepackage{nicefrac}       
\usepackage{microtype}      
\usepackage{xcolor}         
\usepackage{bm}
\usepackage{graphicx}
\usepackage{amsmath, amssymb, bm}
\usepackage{algorithm}
\usepackage{algorithmicx}
\usepackage{algpseudocode}

\title{\LARGE \bf
A Biophysical-Model-Informed Source Separation Framework For EMG Decomposition
}

\author{D. Halatsis (Author), P. Mamidanna (Researcher), J. Pereira (Researcher) 
\\
and D. Farina (Researcher), \textit{Fellow, IEEE} 
\thanks{This project was supported by UK Research and Innovation [UKRI Centre for Doctoral Training in AI for Healthcare grant number EP/S023283/1], the Imperial-META Wearable Neural Interfaces Research Centre and the Onassis Foundation under Scholarship ID: F ZT 012-1/2023-2024.}}

\begin{document}

\maketitle
\thispagestyle{empty}
\pagestyle{empty}

\begin{abstract}

Recent advances in neural interfacing have enabled significant improvements in human-computer interaction, rehabilitation, and neuromuscular diagnostics. Motor unit (MU) decomposition from surface electromyography (sEMG) is a key technique for extracting neural drive information, but traditional blind source separation (BSS) methods fail to incorporate biophysical constraints, limiting their accuracy and interpretability. In this work, we introduce a novel Biophysical-Model-Informed Source Separation (BMISS) framework, which integrates anatomically accurate forward EMG models into the decomposition process. By leveraging MRI-based anatomical reconstructions and generative modeling, our approach enables direct inversion of a biophysically accurate forward model to estimate both neural drive and motor neuron properties in an unsupervised manner. Empirical validation in a controlled simulated setting demonstrates that BMISS achieves higher fidelity motor unit estimation while significantly reducing computational cost compared to traditional methods. This framework paves the way for non-invasive, personalized neuromuscular assessments, with potential applications in clinical diagnostics, prosthetic control, and neurorehabilitation.\newline

\end{abstract}

\section{INTRODUCTION}

Recent advances in neural interfacing have enabled groundbreaking control paradigms for human-computer interaction, augmentation, and rehabilitation \cite{ ctrl_labs}. A fundamental approach in this domain involves interfacing with spinal motor neurons (MNs), the final layer of the motor system responsible for transmitting neural commands to muscles. Unlike other neural signals, which are often weak and noisy, motor neuron activity benefits from a natural amplification process via a vast number of muscle fibers. This amplified signal can be non-invasively decoded from surface electromyography (sEMG) using algorithms such as blind source separation(BSS)—a process commonly referred to as motor unit (MU) decomposition \cite{dario-review, original-ckc, negro2016multi, chen-ica}.

While motor unit decomposition provides precise insights into neural activity, it does not inherently reveal the biophysical properties of individual MNs . These properties serve as critical biomarkers for neuromuscular conditions such as amyotrophic lateral sclerosis (ALS). Despite their importance, most traditional methods for extracting MN properties rely on invasive techniques, limiting their practicality for routine use. This gap highlights the need for an alternative approach that can infer motor neuron properties non-invasively.

Recent advancements in forward EMG generation models offer a potential solution to this challenge. Forward EMG generation models have significantly improved in precision, now capable of accommodating dynamic movement \cite{biomime, halatsis2024modelling}. More specifically, MRI-based anatomical models enable ultra-realistic, personalized forearm simulations, which, when paired with biophysically accurate EMG simulators, can produce highly detailed representations of neuromuscular activity \cite{neurodec}. Can we transition from the traditional BSS framework to a Biophysical-Model-Informed approach? More specifically, how can anatomically accurate forward EMG models improve MU decomposition and enable the estimation of MN properties?

Accurate physical models have long been used to extract information and train agents for various real-world tasks \cite{mamidanna2024inferring}. These models are often leveraged either by training in a simulated environment and transferring the learned knowledge to real-world applications \cite{zhao2020sim} or by using synthetic data generated from forward models to approximate posterior distributions of parameters \cite{cranmer2020frontier}. Meanwhile, physics-informed machine learning (PIML) \cite{raissi2019physics} has driven significant advances in both forward and inverse physical modeling. This suggests an opportunity to explore EMG decomposition from a physical-model inversion perspective.

This approach relies on two key assumptions: first, that physical models will continue to become more accurate and computationally efficient; and second, that the inversion of physical models will become increasingly well-posed and easier as physical modeling transitions from static methods, such as finite element analysis, to more flexible approaches like physics-informed neural networks and neural surrogates. These advancements could pave the way for a more robust, model-based framework for EMG decomposition.

In this work, we propose a novel framework for motor unit decomposition by directly inverting an anatomically and physically accurate forward model in an unsupervised manner. Our key contributions include:
\begin{itemize}
    \item \textbf{A new decomposition framework} that incorporates anatomical information, such as MRI-based modeling, for enhanced accuracy.
    \item \textbf{A simple yet effective model inversion algorithm} that operates without supervision or prior learning.
    \item \textbf{A novel approach for estimating motor neuron }properties directly from sEMG signals.
    \item \textbf{Empirical validation of our framework } in a simulated controlled setting.
\end{itemize}

This framework paves the way for non-invasive, personalized neuromuscular assessments with potential applications in both clinical and assistive technologies.
\section{PRELIMINARIES}

\subsection{EMG Foward  Model}

\begin{figure*}
  \label{forward_model}
  \centering
  \includegraphics[width=1.\linewidth]{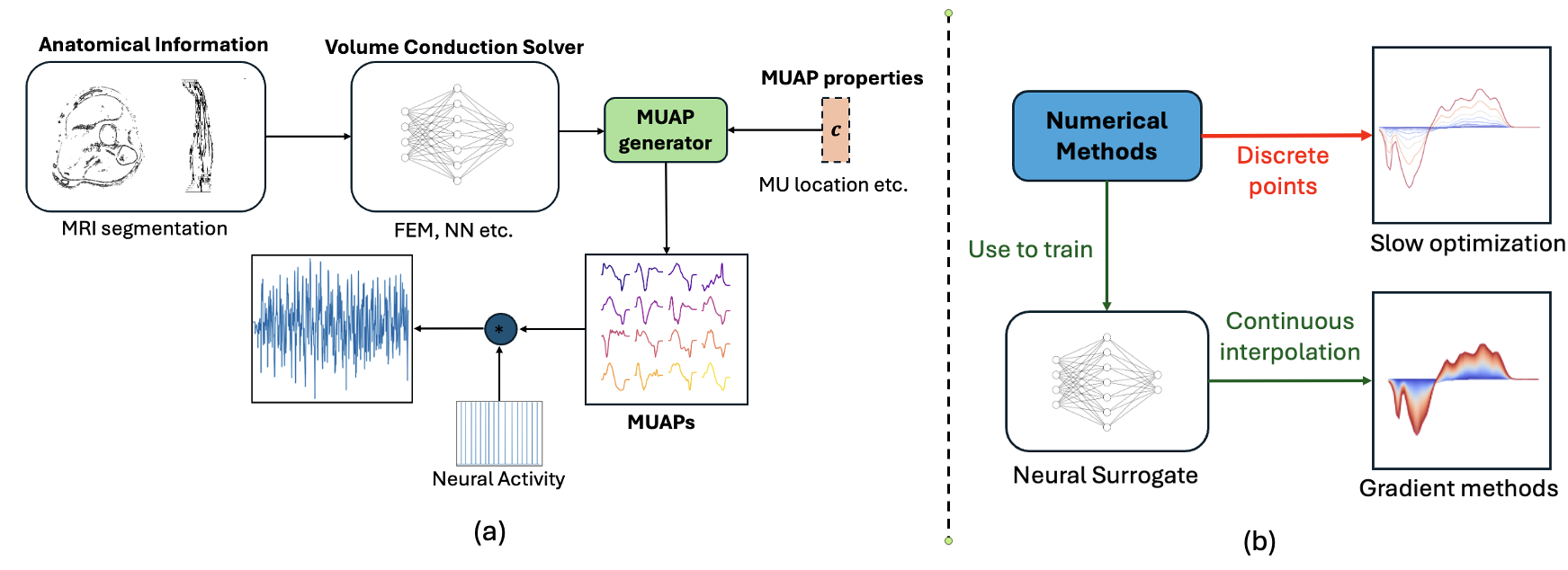}
  \vspace{-1em}
  \caption{\textit{Left}: Simulated EMG forward model overview from volume conductor MRI to MUAPs to EMG.\textit{Right}: Instead of performing costly discrete library-based search, instead train a continuous MUAP generative model to interpolate across samples, that enables the use of gradient based methods}
    \vspace{-1em}
\end{figure*}

Surface electromyography  measures the electrical signals detected on the skin, which result from the activity of motor neurons (MNs) and the muscle fibers they control, collectively forming a motor unit  \cite{merletti2016surface}. When MNs transmit neural signals as spike trains, they trigger action potentials at the innervation zone (IZ)—the site where muscle fibers are initially activated \cite{neural-drive}. These action potentials then propagate longitudinally along the muscle fibers in both directions \cite{farina2004surface}.

While motor neuron firing can be modeled as a spike train, it is not directly detectable on the skin due to the volume conduction effect of biological tissues, which act as a low-pass filter \cite{merletti2016surface}. As a result, the observable signal at the surface consists of the motor unit action potential (MUAP)—the waveform produced by the cumulative effect of action potentials from all fibers within a single MU, after being shaped by tissue conductivity and filtering effects. The MUAP waveform is therefore determined by the anatomical properties of the surrounding tissues, the location of the detector and the intrinsic properties of the motor unit \cite{dario-review}.

\subsection{EMG Data Model and Decomposition}
EMG can be then described as a linear convolutive mixture \cite{original-ckc} of:
\begin{itemize}
    \item \textbf{MUAPs}: which encode both anatomical and motor unit-specific information.
    \item \textbf{Neural drive}, which conveys the firing activity of the motor neurons.
\end{itemize}

Under the assumption of stationary conditions, given $M$ channels and $N$ motor units, EMG can be represented for each channel as:

\begin{equation}
y_{m}[k] = \sum_{n=1}^{N} \sum_{l=0}^{L-1} h_{mn}[k-l]t_{n}[k] + z_{m}[k]\;,
\end{equation}
where $h_{mn}[k]$ is a finite impulse response filter of time support $L$ that corresponds to the MUAP shape, $t_{n}(k)=\sum_{k_{f} \in \Psi_{n}} \delta[k-k_{f}]$ is the delta spike train representing the firings of motor unit $n$ and $\Psi_{n}$ is the set of discharge times of motor unit $n$, and $z_{m}[k]$ is additive i.i.d. noise. This model assumes that the average number of samples between consecutive spikes is significantly larger than L.

To linearize the convolution formulation, the same problem can be represented via its extended matrix representation \cite{original-ckc}. First, we extend the observations, noise and spike train vectors to obtain the following representation:


\begin{equation}
\mathbf{\bar{y}}[k] = \mathbf{H}\mathbf{\bar{t}}[k] + \mathbf{\bar{z}}[k] \;.
\end{equation}
Given $T_{s}$ samples across over time, the multichannel EMG can be hollistically represented as $\mathbf{Y} = \mathbf{HT + Z}$, where $\mathbf{Y} \in \mathbb{R}^{MR \times T_{s}}$, $\mathbf{H} \in \mathbb{R}^{MR \times N(L+R-2)}$, $\mathbf{T} \in \mathbb{R}^{N(L+R-2) \times T_{s}}$, and $\mathbf{Z} \in \mathbb{R}^{MR \times T_{s}}$. In this representation, $R$ is the extension factor, and represents the number of delayed repetitions included to improve condition number of the system. The blind source separation problem involves the estimation of $\mathbf{T}$ given $\mathbf{Y}$ while $\mathbf{H}$ remains unknown. This involves the estimation of the separation matrix $\mathbf{B}$ that approximately inverts $\mathbf{H}$, such that:
\begin{equation}
\mathbf{BY} = \mathbf{BHT} + \mathbf{BZ} \approx \mathbf{T} \;.
\end{equation}


$\mathbf{B}$ contains all unique separation vectors and is referred to as the separation matrix. From this formulation, different optimization procedures can be used to estimate sources, such as fastICA \cite{chen-ica} or convolution kernel compensation (CKC) \cite{original-ckc}, both iteratively estimating one source at a time.

\subsection{Optimal Separation Matrix}
As in \cite{original-ckc}, given the sources, we can estimate the cross-correlation between extended observations and sources as:

\begin{equation}
\mathbf{c}_{\bar{y}, \bar{t}_{n}} = \lim_{card(\Psi_{n}) \rightarrow +\infty} \frac{1}{card(\Psi_{n})} \sum_{k_{f} \in \Psi_{n}} \mathbf{\bar{y}}[k_{f}]\;,
\end{equation}
which is also the spike-triggered average (STA) estimator \cite{STA} of a window of length $R$ from the MUAP at every channel. If we have the ground-truth MUAPs, we generate $\mathbf{c^{*}}_{\bar{y}, \bar{t}_{n}}$ with the central $R$ samples of the MUAP for all channels for each given motor unit. This can be used as the LMMSE optimal estimator of the sources:

\begin{equation}
\mathbf{\hat{\bar{t}}}_{n}[k] = \mathbf{c^{*}}_{\bar{y}, \bar{t}_{n}}^{T} \mathbf{C}_{\bar{y}\bar{y}}^{-1} \mathbf{\bar{y}}[k] \;,
\end{equation}
where $\mathbf{C}_{\bar{y}\bar{y}}^{-1}$, the covariance matrix of the extended observations. 





\subsection{Biophysical Modelling}

A key challenge in both forward EMG generation and motor unit decomposition is obtaining the MUAP waveforms for a given anatomical structure. Once the MUAP waveforms are available, generating or decomposing EMG signals becomes a straightforward task. The computation of MUAPs requires solving the electric potential propagation in biological tissue, which can be described as a quasi static Maxwell equation and formulated using the Poisson equation with Neumann boundary conditions \cite{farina2004surface}:
\begin{align}
    \nabla \cdot (\bm{\sigma} \nabla v ) &= -I \quad \text{in } \bm{\Omega}  \\
     \nabla v \cdot \bm{n} &= 0 \quad \text{on } \partial\bm{\Omega} \;,
\end{align}
where $\bm{\Omega}$ is the domain of definition of the three-dimensional volume conductor with boundary  $\partial\bm{\Omega}$, $v$ is the electric field potential, $\bm{\sigma}$ the conductivity tensor and $I$ is the current source term.

For simple symmetrical volume conductors, analytical methods can be applied to solve this equation \cite{farina2004surface}. However, for more complex and realistic anatomical topologies, numerical methods such as finite element methods (FEM) or neural solvers are required \cite{neurodec, halatsis2024modelling}. The resulting solution field, combined with a set of motor unit properties, enables the generation of MUAPs for the modeled volume conductor.


\textit{BioMime}\cite{biomime} is a conditional generative model designed to replicate the MUAPs produced by a high-fidelity numerical model . It is built upon personalized anatomical data from a single subject and generates MUAPs for motor units in eight superficial muscles of the forearm. The model employs a $10 \times 32$ electrode grid positioned around the distal third of the forearm.

The key input parameters that determine the MUAP shape in Biomime are:

\begin{itemize}
    \item \textbf{Motor Unit Location}: The spatial position of the motor unit within the muscle in polar coordinates. It consists of MU depth and medial-lateral angle.
    \item \textbf{MU size}: The total number of muscle fibers innervated by the motor neuron, effectively defining the motor unit size.  
    \item \textbf{Innervation Zone}: The region along the muscle fibers where the motor neuron establishes synaptic connections.  
    \item \textbf{Fiber Length}: The proportional length of muscle fibers relative to the tendon.  
    \item \textbf{Conduction Velocity}: The speed at which action potentials propagate along muscle fibers.  

\end{itemize}

By leveraging a neural-network-based generative approach \cite{goodfellow_gans}, Biomime efficiently produces MUAPs across a wide range of motor unit properties without the computational burden of traditional forward models. This capability is particularly advantageous for inverse modeling applications and motor unit property estimation.

\section{MATERIALS AND METHODS}

Our objective is to develop a method that leverages anatomical information, such as MRI-derived forearm structures, to allow a biophysical-model-informed EMG decomposition and simultaneously estimate motor unit properties. Conceptually, this can be viewed as performing decomposition while also solving the inverse problem—mapping a given set of MUAPs back to the underlying physiological conditions that generated them—in a single step.


\subsection{Naive library of MUAPs approach}
A straightforward initial approach to this problem would involve generating a comprehensive MUAP library across a sufficiently diverse range of conditions and then performing a discrete search using an optimization algorithm. By running a large number of simulations, we can construct a library of candidate MUAP waveform sets, denoted as $\mathcal{S}_{lib} = \{\bm{m_{c_1}}, \bm{m}_{c_2}, \dots, \bm{m}_{c_{\nu}}\}$, where each $\bm{m_{c_i}}$ represents a set of MUAP waveforms recorded across multiple electrodes, generated under a specific set of initial conditions $c_i$.

Using the linear minimum mean square error (LMMSE) optimal estimator to transform MUAPs $\bm{m_{c_i}}$ into a separation vector $\bm{r}_{\bar{y}}(\bm{m_{c_i}})$ for decomposition, we can formally define the problem as:

\begin{align}
    \text{find} \;\bm{m_{c_i}}:  \kappa(\bm{r}_{\bar{y}}(\bm{m_{c_i}})^{T} \mathbf{C}_{\bar{y}\bar{y}}^{-1} \mathbf{\bar{y}}) \geq \text{threshold}\;,
\end{align}
where $\kappa$ is a measure of non-Gaussianity, such as kurtosis or estimated negentropy, and the threshold determines the minimum acceptable value for identifying a valid source.

While high-fidelity numerical models can generate extensive MUAP libraries, this brute-force search approach is computationally infeasible due to the sheer volume of required simulations.

\begin{figure*}
  \label{biomime_decomp_main_diagram}
  \centering
  \includegraphics[width=0.7\linewidth, trim={0 1cm 0 0}]{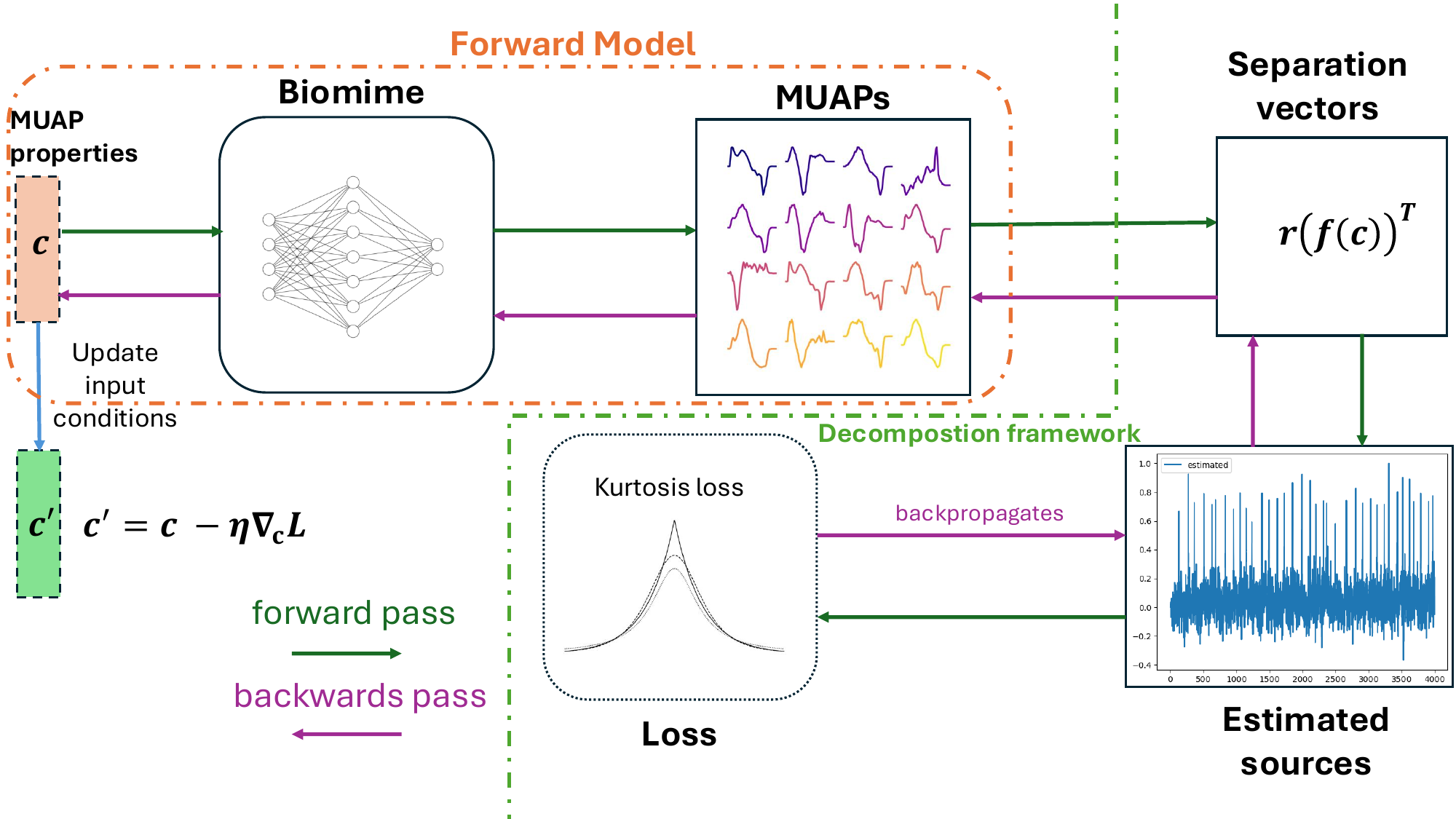}
  \caption{ Full decomposition framework. The framework attempts to optimize the input parameters $\bm{c}$ to maximize kurtosis on the separated sources. In the forward pass the model generate MUAPs, to separation vectors to sources. The backward pass involves backpropagating throughout our entire framework.}  
\end{figure*}

\subsection{Gradient-Based Continuous Optimization of Conditions}

Instead of relying on discrete search, we can reformulate the problem by assuming that the MUAP library is represented by a continuous and differentiable function mapping the initial condition space $D_{\bm{c}}$ to the MUAP space $D_{\bm{m}}$:

\begin{align}
    \bm{m}_{\bm{c}_i} = f(\bm{c}_i) , \;\text{where} \; f: D_c \rightarrow D_m.
\end{align}
We can construct such a function by training a neural network on a set of MUAP samples. Neural networks are highly general function approximators \cite{universal-approximator}, with advantageous properties such as continuity and differentiability. By parameterizing the MUAP waveforms using a neural network $f$, we establish a well-defined continuous optimization landscape, allowing gradient-based methods \cite{lecun1998gradient} to efficiently solve the following optimization problem:

\begin{align}\label{decomp_main}
    \arg \max_{\bm{c}_i } : \kappa(\bm{r}_{\bar{y}}(f({{\bm{c}_i}}))^{T} \mathbf{C}_{\bar{y}\bar{y}}^{-1} \mathbf{\bar{y}}).
\end{align}

This formulation corresponds to identifying the optimal forward conditions that maximize the non-Gaussianity of the estimated sources \cite{negro2016multi, original-ckc}.
\\
\\
\textbf{Motor Neuron Firing and Properties}. By solving the optimization problem in Eq. \ref{decomp_main}, we obtain both the neural drive of the MNs from the separated sources and the corresponding motor neuron properties $\bm{c}$.
\\
\\
\textbf{Lower dimension optimization.} In traditional convolutive BSS \cite{original-ckc, chen-ica}, the optimization domain resides in $\mathbb{R}^{MR}$, where $M$ represents the number of EMG channels and $R$ is the extension factor. However, in Eq. \ref{decomp_main}, the optimization occurs in a significantly lower-dimensional space $\mathbb{R}^{K}$, where $K$ is the dimensionality of the motor unit property vector $\bm{c}$.
\\
\\
\textbf{Constrained optimization:} Since the optimization is performed directly on motor unit properties, it can be constrained to specific subsets of the domain to promote the uniqueness of sources. For instance, constraints can be applied to identify MUAPs of a particular size or within specific regions of the muscle. This property is particularly beneficial for EMG decomposition, as it mitigates the need for performing multiple decompositions using the peel-off method.

\subsection{Practical Challenges}

\begin{figure*}
  \label{heatmaps}
  \centering
  \includegraphics[width=1.\linewidth, trim={0 6cm 0 0}, clip]{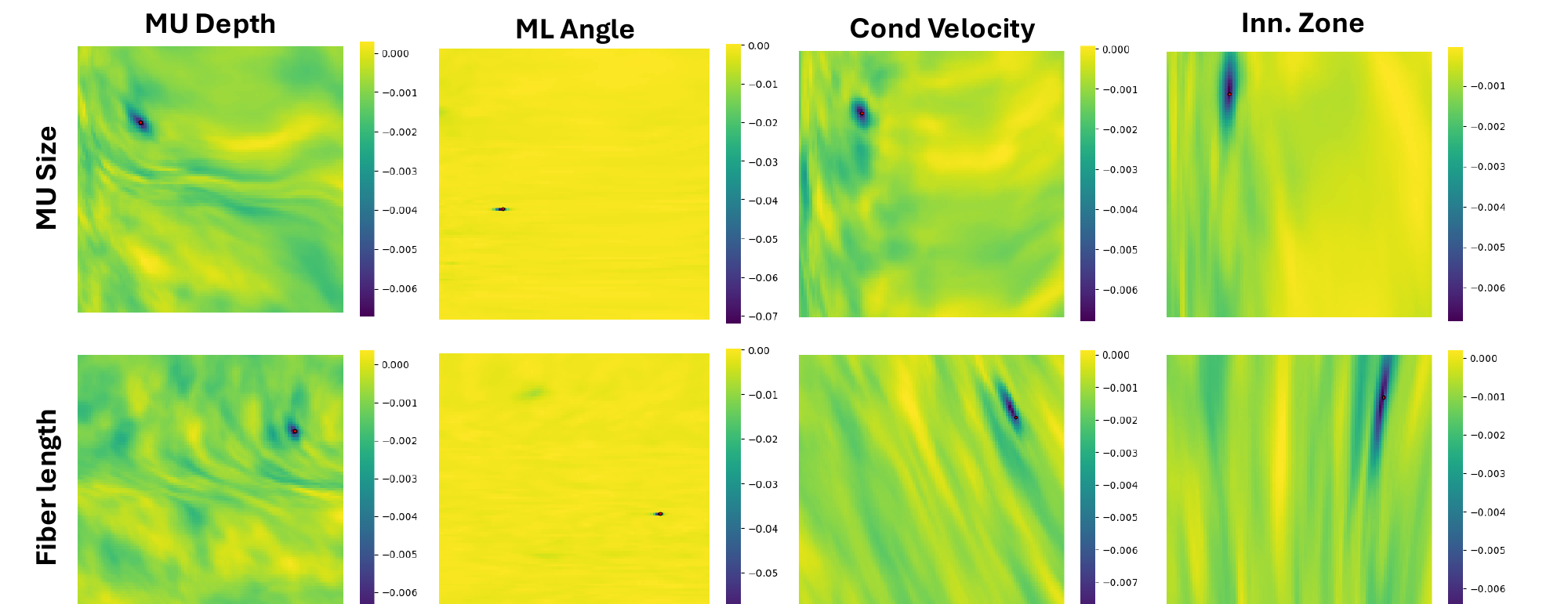}
  \caption{2D slice of the true optimization landscape across different input variables. Heatmaps were generated by evaluating a grid of the loss function (negative kurtosis) over the displayed parameters while holding all other parameters at their true values. The true condition is marked with a red circle. Although the optimization landscape exhibits extended flat regions and highly non-convex features, a distinct local minimum is clearly visible.
}
\end{figure*}

In our work, we chose \textit{Biomime} as the base for our MUAP generator. Biomime is a conditional variational autoencoder (VAE) \cite{kingma2013auto} trained with adversarial learning \cite{goodfellow_gans}. However, since our focus is solely on its deterministic MUAP generation capabilities, we eliminate its stochasticity by removing the latent Gaussian variable $\bm{z}$ associated with the VAE. This transformation results in a deterministic function that directly maps input conditions to MUAPs. The final input to our model consists of the six MUAP conditions previously discussed. The full framework can be viewed in \ref{biomime_decomp_main_diagram}
\\
\\
\textbf{Model Invertibility.} Our optimization objective in Eq. \ref{decomp_main} is to perform single-step decomposition and model inversion by minimizing with respect to the input x. However, since neural networks are not inherently bijective, invertibility is not guaranteed, leading to two key challenges. First, multiple sets of input conditions may produce identical MUAPs, introducing ambiguity in the inversion process. Second, over-parameterized networks can create redundancies, making the optimization ill-posed, unstable, and difficult to solve efficiently \cite{ papamakarios2021normalizing}.
\\
\\
\textbf{Challenges in Source Separation Optimization}. Model inversion becomes even more complex when coupled with decomposition. Traditional decomposition algorithms typically rely on fixed-point iteration \cite{hyvarinen1999fast} rather than first-order gradient methods.

The optimization landscape shaped by non-Gaussianity loss often contains large flat regions, making it difficult to navigate with fixed-learning-rate gradient methods \cite{hyvarinen2000independent}. Since our framework does not support fixed-point iteration, alternative strategies are needed to ensure convergence to the correct initial conditions.

Moreover, blind source separation problems inherently admit multiple valid solutions \cite{hyvarinen2000independent}. In conventional BSS, a single source can correspond to multiple separation vectors, and multiple sources can produce identical observed signals. Techniques like orthogonalization \cite{negro2016multi} and the peel-off method are often used to enforce uniqueness. Even if we assume Biomime is bijective and fully invertible, the optimization landscape still permits multiple valid solutions.
\\
\\
\textbf{Visualizing the Optimization Landscape.} Empirical observations suggest that relying solely on gradient-based methods is ineffective, as the optimizer often gets trapped in flat regions or local minima. To illustrate this, we sample and plot the 2D optimization landscape across two of the six parameters for a specific MUAP (see Section IV).

These heatmaps represent 2D slices of a six-dimensional optimization space, revealing large flat regions where fixed learning rates struggle. Despite the landscape's overall non-convexity, we observe that in localized low-dimensional settings, a single local minimum corresponds to the true parameters. While the full optimization landscape may contain many local minima and flat regions, a localized view near the correct values confirms that the true source aligns with the local minimum

\subsection{Hybrid optimization algorithm}
To address these challenges, we employ a hybrid optimization algorithm, as outlined in Algorithm \ref{alg:decomposition_alg}. Our approach leverages the fact that the forward pass and loss computation are significantly less expensive than backpropagation.

Instead of relying solely on gradient-based updates, our algorithm begins by evaluating a large number of candidate points using a computationally inexpensive forward pass (Steps 4-7). From these, a subset of promising candidates is selected (Step 7), and gradient descent is applied only to this smaller batch (Steps 8-9). 

To prevent converging to duplicate sources, Step 5 ensures that new regions are explored for gradient descent, increasing the likelihood of discovering novel sources. Overall, the sampling strategy helps mitigate issues associated with flat regions and local minima by initializing optimization from more favorable starting points. Additionally, it encourages convergence toward distinct sources.


\begin{algorithm}
\caption{Neural Forward Model EMG Decomposition}
\label{alg:decomposition_alg}
\begin{algorithmic}[1]
    \State \textbf{Input:} Neural MUAP Generator $f$, motor units target $N$, MU search size $K$, batch size $B$, condition space $D_c$, Previous solution similarity threshold $\epsilon$, learning rate $\eta$.
    \State \textbf{Initialize:} $c_{\text{sources}} \gets \emptyset$
    \While{$|c_{\text{sources}}| < N$}
        \State sample $K$ points $\bm{c}_{\text{guess}}$ from $D_c$
        \State $\bm{c}_{\text{guess}} \leftarrow \{\bm{c} \in \bm{c}_{\text{guess}}: \forall \bm{c'} \in \bm{c}_{\text{csources}} : ||\bm{c}-\bm{c'}||_2 < \epsilon  \}$
        \State $\mathcal{L}(f(\bm{c}_{\text{guess}})) \leftarrow -\kappa(\bm{r}_{\bar{y}}(f({{\bm{c}_\text{cand}}})) ^{T}\mathbf{C}_{\bar{y}\bar{y}}^{-1} \mathbf{\bar{y}})$ 
        \State $\bm{c}_{\text{cand}} \leftarrow B$ points with lowest loss from $\bm{c_\text{guess}}$   
        \While{${\text{has\_not\_converged}}$}
            \State $\mathcal{L} \leftarrow -\kappa(\bm{r}_{\bar{y}}(f({{\bm{c}_\text{cand}}})) ^{T}\mathbf{C}_{\bar{y}\bar{y}}^{-1} \mathbf{\bar{y}})$ 
            \State \quad $\bm{c}_{\text{cand}} \gets \bm{c}_{\text{cand}} - \eta \bm{\nabla}_{\bm{c}} \mathcal{L}$
        \EndWhile
        \State prune $ \bm{c} \in \bm{c}_\text{cand}: -\kappa(\bm{r}_{\bar{y}}(f({{\bm{c}}}))^{T}\mathbf{C}_{\bar{y}\bar{y}}^{-1} \mathbf{\bar{y}}) > \text{threshold}$
        \State $\bm{c}_\text{sources} \leftarrow \bm{c}_\text{sources} \cup \bm{c}_\text{cand}$
    \EndWhile
    \State $\bm{s} \leftarrow \bm{r}_{\bar{y}}(f({{\bm{c}_\text{cand}}}))^{T}\mathbf{C}_{\bar{y}\bar{y}}^{-1} \mathbf{\bar{y}}$
    \State \textbf{Output:} MU properties $\bm{c}_{\text{sources}}$, Separated sources $\bm{s}$
\end{algorithmic}
\end{algorithm}

\section{EXPERIMENTAL VALIDATION}

We experimentally validated the feasibility of our framework in a controlled simulated setting where the ground truth was known. To achieve this, we generated training data using our existing simulation pipeline:

\begin{itemize}
    \item \textbf{Motor Unit Properties:} We used Biomime to generate MUAPs for \textbf{100 motor units}, with their properties randomly sampled according to the power-law distribution of motor unit sizes \cite{merletti2016surface}.
    \item \textbf{Neural Activation:} For each motor neuron (MN), we generated a random spike train to represent source activation, with a mean inter-spike interval of $100$ ms and a standard deviation of $30$ ms.
\end{itemize}  

These simulated conditions served as the ground truth. Our model, initialized with random conditions, attempted to perform source separation on the EMG signals obtained from their convolutive mixture. To evaluate robustness, we introduced varying levels of noise into the generated EMG and analyzed the model’s performance under different conditions.

The performance of our method was evaluated based on the accuracy of the EMG decomposition, i.e. how many sources were accurately decomposed, and our other metric is the how accurate the estimated MU properties compared to the true ones.

To quantify these objectives, we define the following evaluation metrics:
\begin{itemize}
    \item \textbf{Number of MU retrieved (\#MU):} The number of distinct MUs detected.
    \item \textbf{Average Silhouette Score (SIL):} The mean silhouette coefficient computed from binary k-means clustering of the estimated sources. This metric includes both accepted and rejected sources.  
    \item \textbf{Ratio of Rejected MUAPs (disc ratio):} The percentage of sources discarded after the optimization process is completed. This metric quantifies the proportion of points that fail to converge to a valid source.  
    \item \textbf{Condition RMSE (c RMSE):} The RMSE between the true MUAP properties and the estimated ones.
    \item \textbf{Condition Accurary (c Acc)}: We consider a condition to be true if the $MSE < 0.01$.
\end{itemize}

\begin{table}[h]
    \caption{Performance metrics for different batch sizes and SNR values.}

    \label{main_results_Table}
    \centering
    \setlength{\tabcolsep}{3pt} 
    \renewcommand{\arraystretch}{0.9} 
    \begin{tabular}{lccccc}
        \toprule
        batch size and SNR& \small \textbf{\#MU }& \small \textbf{c RMSE} & \small \textbf{c Acc} & \small  \textbf{SIL} & \small \textbf{disc ratio} \\
        \midrule
        \textbf{baseline} & 5 & - & - & 0.91325 & - \\
        bsize=20 & 61.2($\pm$3.5) & 0.0044 & 1.0000 & 0.9073 & 0.0450 \\
        bsize=40 & 74.3($\pm$2.8) & 0.0019 & 1.0000 & 0.9008 & 0.0375 \\
       \textbf{bsize=60} & \textbf{85.8}($\pm$2.2) & \textbf{0.0106} & \textbf{0.997} & 0.8865 & 0.0383 \\

        bsize=20, SNR=10 & 39.3($\pm$1.7) & 0.0498 & 0.9695 & 0.7670 & 0.6550 \\
        bsize=20, SNR=5 & 24.5($\pm$1.4) & 0.0737 & 0.5882 & 0.6205 & 0.4250 \\
        \bottomrule
    \end{tabular}
    \label{tab:results}
\end{table}

\begin{figure}
  \label{heatmaps}
  \centering
  \includegraphics[width=1.\linewidth, trim={0 0.7cm 0 0}, clip]{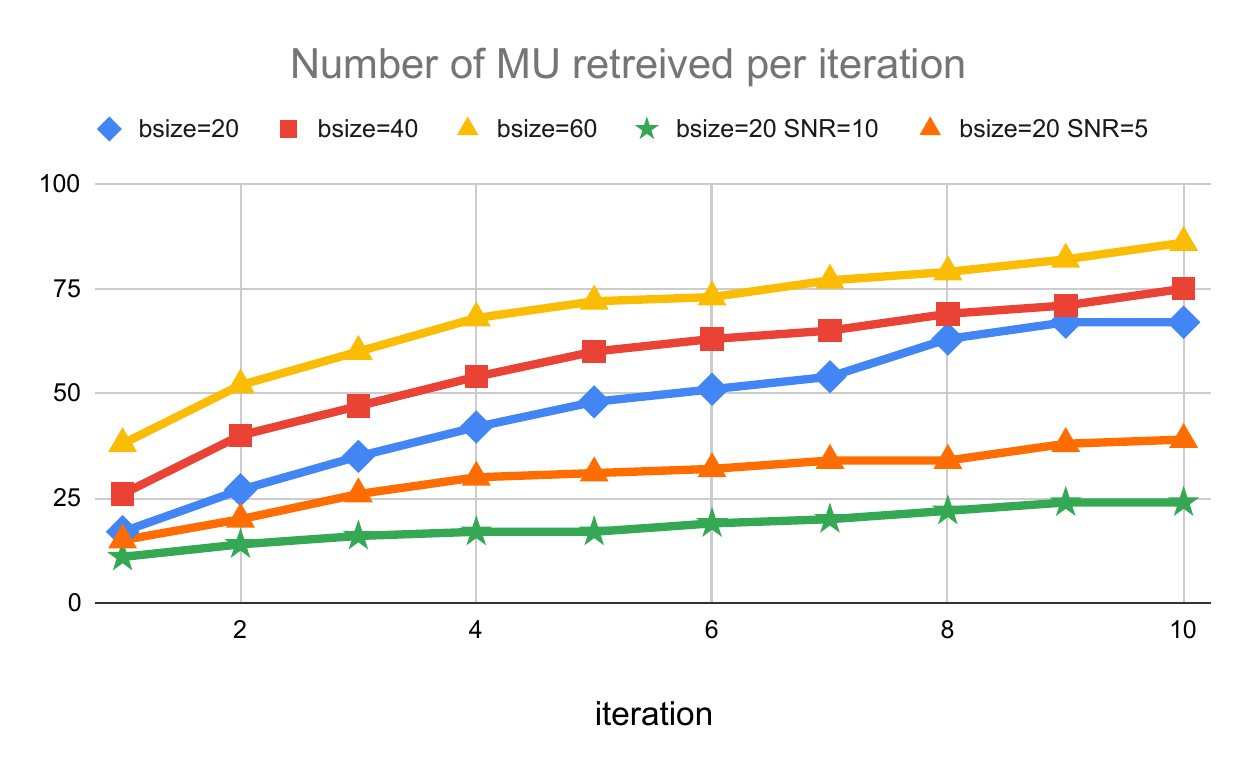}
  \vspace{-1.5em}
  \caption{Cumulative number of motor units retrieved per iteration. As observed, using a larger batch size increases the number of points optimized via gradient descent, enabling the retrieval of a greater proportion of the total motor units.}
\vspace{-1em}
\end{figure}

\textbf{Framework Setup:} We choose negative kurtosis as our loss function $\kappa(x) = x^{4}$. We initialized the probe size to $K = 10000$, which accounted for approximately one-third of the total decomposition time, and set the batch size to 100 points, ensuring compatibility with main memory constraints. To simplify the experiments, we did not impose any similarity threshold on previously identified solutions (i.e., $\epsilon = \infty$). A source was accepted or rejected based on the criterion in line 12 of Algorithm \ref{alg:decomposition_alg}, where a source was considered valid if its calculated silhouette score exceeded $SIL > 0.9$.

For optimization, we employed the Adam optimizer \cite{kingma2014adam}, running for 100 epochs per iteration with an initial learning rate of $\eta = 10^{-2}$, which decayed to $\eta = 10^{-3}$ for the final 30 epochs. The framework required approximately 90 seconds per iteration on an NVIDIA RTX 2080Ti, demonstrating a competitive speed compared to traditional decomposition methods that utilize the peel-off technique. 

To ensure robustness, we averaged our experimental results over five independent repetitions. As a baseline, we also evaluated a standard FastICA implementation without peel-off, which identified an average of only five unique sources out of 100 decomposed ones.

\subsection{Source separation results}

Overall, our method effectively retrieves the majority of motor units (MUs) within a few iterations while maintaining high accuracy in the extracted MU properties. Notably, our approach consistently recovers between 60 and 80 MUs directly from the EMG, without requiring the peel-off method or imposing any similarity threshold on previously identified solutions ($\epsilon$).  

From Table \ref{main_results_Table}, we observe that using a batch size of 60 MUs per iteration in Algorithm \ref{alg:decomposition_alg}, our method successfully retrieves on average \textbf{86 out of 100 MUs} within just 10 iterations. While reducing the batch size can improve computational efficiency, the optimal value should generally be chosen based on available memory constraints.  

Examining the source discard ratio, we find that the algorithm almost always converges to a valid source. However, as the iterations progress, fewer of these sources remain unique. This behavior reflects a previously discussed challenge: the optimization landscape contains multiple correct local minima, meaning that many distinct sets of conditions can lead to the correct estimation of a source. In practice, our algorithm tends to converge more frequently to specific local minima, even with randomized initialization, while certain minima—likely corresponding to very small or deep motor units—appear to be much harder to reach.  

\subsection{Motor Unit Property Results}  

The retrieved motor unit conditions demonstrate high accuracy, with a mean squared error (MSE) as low as $10^{-2}$, corresponding to an approximate relative error of $3\%$ between the estimated and true parameter values. However, in the run with a batch size of 60, this error increases as more MUs are identified, reaching an MSE of $0.01$, which translates to an average relative error of $10\%$. This suggests that as the algorithm uncovers previously inaccessible MUs, the range of acceptable conditions broadens, leading to slightly less precise properties estimates.  

Despite this, the overall results strongly indicate that Biomime is nearly invertible. Instances where different initial conditions produce MUAPs similar enough to enable correct EMG decomposition are exceedingly rare, reinforcing the reliability of the approach.  

\subsection{Effect of Added Noise}  

The accuracy picture changes significantly in the presence of added noise. Under noisy conditions, the RMSE increases by factors of 40 and 80, respectively, compared to the noiseless case. This degradation is likely due to the fact that low-amplitude MUAPs are disproportionately affected by noise, resulting in a more diffused optimization landscape where minima become shallower and broader. Consequently, accuracy is reduced, and in many cases, the original source information is completely lost.  

\section{DISCUSSION}
The advancement of biophysical modeling and neural networks has enabled both highly accurate simulations and efficient model inversion, facilitating sensitivity analysis of complex systems. Building on these capabilities, our goal is to transition from traditional BSS to a biophysical-model-informed source separation framework.

This work represents an initial step toward that goal by integrating neural forward models, source separation techniques, and neural network inversion methods into a unified framework. Our empirical evaluation in a simulated setting demonstrates strong evidence that this approach is viable, showing promising performance in decomposing EMG signals. As a next step, we plan to validate its feasibility on real EMG data.

More broadly, this work aims to establish a foundation for further innovation. By demonstrating the advantages of a biophysical-model-informed approach, we hope to inspire the development of more advanced \textbf{dynamic forward models capable} of handling a wider range of conditions,\textbf{ universal models} that can adapt to individual users, and \textbf{hybrid source separation techniques} that integrate both traditional and model-based approaches. We anticipate that frameworks like this will contribute to the broader evolution of neuromuscular signal processing and foster deeper connections between biophysical modeling and EMG decomposition.




\bibliographystyle{IEEEtran}  
\bibliography{references.bib}

\end{document}